\documentclass[floats,aps,showpacs,twocolumn]{revtex4}

\usepackage[dvips]{graphicx}
\usepackage{amsfonts}
\usepackage{amsmath,amssymb}
\usepackage{dsfont}
\usepackage{placeins}
\usepackage{afterpage}
\usepackage{flafter}

\usepackage{color}
\usepackage{cancel}

\begin{document}

\newcommand{\iexp}{\text{i}}
\newcommand{\ui}{\text{i}}
\newcommand{\diff}{\text{d}}
\newcommand{\Rac}{ R^{\text{ac}} }
\newcommand{\heff}{\hbar_{\text{eff}}}

\title{A switching mechanism in periodically driven quantum systems
with dissipation}

\author{Roland Ketzmerick}
\author{Waltraut Wustmann}
\affiliation{Institut f\"ur Theoretische Physik,
            Technische Universit\"at
            Dresden, 01062 Dresden, Germany}

\date{\today}

\begin{abstract}
We introduce a switching mechanism in the asymptotic occupations
of quantum states induced by the combined effects of a periodic driving
and a weak coupling to a heat bath.
It exploits one of the ubiquitous avoided crossings in driven systems
and works even if both involved Floquet states have small occupations.
It is independent of the initial state and the duration of the driving.
As a specific example of this general switching mechanism we show how an
asymmetric double well potential can be switched between the
lower and the upper well
by a periodic driving that is much weaker than the asymmetry.
\end{abstract}


\maketitle

\section{Introduction}

The interplay between a coherent periodic driving force
and the incoherent damping of a thermal environment
enriches the dynamics of a quantum system
and opens new potential applications~\cite{GriHae1998}.
In addition to controlling the transient
dynamics,
e.g.\ with respect to
tunneling~\cite{KohUteHaeDit1998, MakMak1995, SolBur2005},
a control of the asymptotic state is desirable.
The ability to design a system's probability distribution, e.g.\
to switch between two macroscopically distinguishable states, in
the presence of a thermal environment is a key to quantum control
techniques.

The paradigmatic model for switching is a double well potential
which is experimentally realized
in superconducting quantum interference devices (SQUIDs)~\cite{MakSchShn2001},
atom-optical potentials~\cite{KieETAL2008, DeuETAL2000},
spin tunneling in condensed matter~\cite{LueDel2003}
or in the transfer of protons along chemical bonds~\cite{DouLahZew1996}.
In some cases the model can be restricted to a two-level system.
Different approaches for switching by a population inversion in driven
two-level systems have been proposed, e.g.\
induced by symmetry-breaking~\cite{AdaWinGriWei1999},
structured environments~\cite{GooThoGri2004}
strong nonequilibrium noise~\cite{GoyHae2005},
or strong driving~\cite{StaDohBar2005}.
However, the restriction to a two- or a three-level system
limits the possible switching mechanisms.

Time-periodic quantum systems are best described by Floquet states,
which are solutions of the Schr\"odinger equation without the coupling
to the environment.
When coupling the time-periodic system weakly to a thermal bath,
all Floquet states are asymptotically populated with occupation probabilities,
which can be determined within a Floquet-Markov
approach~\cite{BluETAL1991, KohDitHae1997, BreHubPet2000, Koh2001, HonKetKoh2009}.
These occupations are quite different from the canonical
distribution of Boltzmann weights in undriven systems
and so far lack an intuitive understanding.

\begin{figure}[b]
\includegraphics[width=\columnwidth]{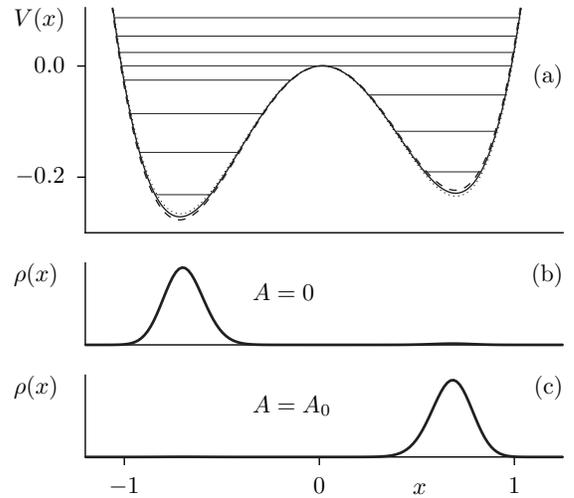}
\caption{
a) Asymmetric double well potential and its eigenenergies
without driving, $A=0$ (solid line),
and the almost indistinguishable variation of the potential
for a small driving amplitude $A_0 \approx 0.008$ (dashed and dotted line).
b) Asymptotic probability density $\rho(x)$
for $A=0$ and a small temperature $1/\beta=1/100$,
with almost all probability in the left well.
c) Cycle-averaged asymptotic probability density $\rho(x)$
according to Eq.~\ref{eq:rho_x} for $A=A_0$,
with more than 99 \% of probability in the right well,
demonstrating a weak driving induced switching to a
macroscopically different state.
See Fig.~\ref{fig:bild2} for parameters.
}
\label{fig:bild1}
\end{figure}

In this paper we demonstrate a dramatic property of
time-periodically driven quantum systems weakly
coupled to the environment:
The asymptotic state can be switched
to an almost orthogonal state by a small parameter variation.
This is in stark contrast to time-independent systems,
where the asymptotic occupations are determined by Boltzmann weights
and vary slowly with a parameter.
The proposed switching mechanism exploits one of the ubiquitous
avoided crossings in driven systems and works even if
both involved Floquet states have small occupations.
As a specific example of this general switching mechanism
we show for an asymmetric double well
potential, see Fig.~\ref{fig:bild1}, that a weak periodic driving
switches the cycle-averaged asymptotic probability density from the
ground state of the undriven system in the left well
to the right well.
Note, that the periodic driving is much weaker than the asymmetry,
see Fig.~\ref{fig:bild1}(a),
and therefore this switching is unrelated to previous studies
on hysteretic switching in a driven dissipative
double well~\cite{ThoJun1997, ThoReiJunFox1998}.
We explain the switching mechanism by an effective rate equation,
which combines the effects of the coherent driving at an
avoided crossing of two Floquet states with the incoherent
bath coupling.

The paper is organized as follows:
In Sec.~\ref{sec:model} our model
for the periodically driven, dissipative double well is introduced.
The switching mechanism is investigated in Sec.~\ref{sec:switching}.
We finally conclude and discuss advantages of the switching mechanism in
Sec.~\ref{sec:conclusion}.

\section{The Model system}\label{sec:model}

As an example we study a particle in an asymmetric double well potential in
the quantum regime, where the ground state is in the left well
and the first excited state is in the right well,
see Fig.~\ref{fig:bild1}(a).
It is driven by an additive time-periodic force,
leading to the system Hamiltonian
\begin{equation}
H_s(t) = \frac{p^2}{2m}
+ V_0 \left[ \frac{x^4}{x_0^4} - \frac{x^2}{x_0^2} 
+ \frac{x}{x_0} \left( \mu  + A \cos \Omega t \right) \right]
,
\end{equation}
where $\mu$ is the asymmetry parameter of the double well potential
and $A$ and $\Omega$ are the driving amplitude and frequency, respectively.
We introduce the dimensionless quantities
$\tilde x = x / x_0$, 
$\tilde H_s=H_s / V_0$, 
$\tilde t = t \cdot V_0/\hbar$,
$\tilde \Omega = \Omega \cdot \hbar/V_0$,
and $\heff = \hbar / (\sqrt{m V_0} x_0)$.
In the following we omit the tilde and then the dimensionless Hamiltonian reads
\begin{equation}
H_s(t) = -\frac{\heff^2}{2} \frac{\partial^2}{\partial x^2} 
+ x^4 - x^2 + x \left( \mu + A \cos \Omega t \right)
.
\end{equation}
The Schr\"odinger equation of a periodically driven quantum system
has according to the Floquet theorem solutions of the form
$\psi_i(t) = e^{-\iexp \varepsilon_i t} u_i(t)$, with $u_i(t+T) = u_i(t)$
and $T=2\pi/\Omega$ the period of the driving.
The time-periodic parts $u_i(t)$ of the Floquet states form a
complete orthonormal set at all times.
The quasienergies $\varepsilon_i$ can be chosen to lie in the
interval $[0,\Omega)$.

The coupling to a heat bath is modeled in a standard way
by a Hamiltonian~\cite{Wei1999}
\begin{equation}\label{eq:Hamiltonian_tot}
H(t) = H_s(t) + H_b + H_{sb}
.
\end{equation}
The bath Hamiltonian
$H_b = \sum_n \left( \frac{p_n^2}{2m_n} + \frac{m_n \omega_n^2}{2}  x_n^2 \right)$
describes an ensemble of noninteracting harmonic oscillators
coupled via $H_{sb} = x \sum_n c_n x_n$ to the system.
The properties of the system-bath coupling are
given in terms of the spectral density of the bath
$J(\omega) := \frac{\pi}{2} \sum_n \frac{c_n^2}{m_n \omega_n}  \left[
\delta \left( \omega - \omega_n \right) - \delta \left( \omega + \omega_n \right) \right]$.
In the continuum limit the spectral density is assumed to be a
smooth function which is linear for an Ohmic bath.
An exponential cutoff beyond the spectral mode $\omega_c$ leads to
$J(\omega) = \eta \omega \, e^{-\left|\omega\right|/\omega_c}$,
where $\eta$ is proportional to the classical damping coefficient.

In the presence of the heat bath the state of the system is described by
the reduced density operator $\rho(t)$.
Its equation of motion for time-periodic quantum systems has been derived
within the Floquet-Markov
approach~\cite{BluETAL1991, KohDitHae1997, BreHubPet2000, Koh2001, HonKetKoh2009}:
Herein the Floquet formalism ensures a non-perturbative treatment
of the driven systems coherent dynamics.
The coupling to the heat bath is treated perturbatively, which is
valid in the limit of weak coupling between the driven system and the bath.
This approximation requires a rapid decay of bath correlations compared
to the typical relaxation time of the system
and we further require $\Omega \ll \omega_c$.
In the following we restrict the discussion to the limit of large times,
larger than the relaxation time.
In this limit the density matrix $\rho_{ij}$ in the basis of the periodic parts
$u_i(t)$ of the Floquet states is approximated as 
time-independent~\cite{KohDitHae1997, HonKetKoh2009}.
Note, that the corresponding density operator,
$\sum_{i,j} | u_i(t) \rangle \rho_{ij} \langle u_j(t) |$,
is time-periodic because of the inherent time-dependence of the $u_i(t)$.
The matrix elements $\rho_{ij}$ obey the rate equation
\begin{eqnarray}\label{eq:DGL_rho}
\lefteqn{\ui \left( \varepsilon_{i} - \varepsilon_{j} \right) \rho_{ij} =} \\
&&  - \sum_{k,l} \Bigl\{
    \rho_{lj} R_{ik;lk} + \rho_{il} R^*_{jk;lk} - \rho_{kl} \Bigl(R_{lj;ki} + R^*_{ki;lj} \Bigr)
\Bigr\} 
. \nonumber
\end{eqnarray}
The complex rates
$R_{ij;kl} = \pi \sum_m x_{ij}(m) x^*_{kl}(m) g(\varepsilon_l-\varepsilon_k - m \Omega)$
describe bath-induced transitions between the Floquet states,
the $x_{ij}(m)$ are the Fourier coefficients of the time-periodic
matrix elements $\langle u_i(t)|x|u_j(t)\rangle$,
and $g(\omega)$
is the correlation function of the bath coupling operator.
The latter is given by
$g(\omega) = n_\beta(\omega) J(\omega)/\pi$
with the spectral density $J(\omega)$ and the thermal occupation
number $n_\beta(\omega)$
of the boson bath with temperature $1/\beta$.
In numerical studies of the rate equation \eqref{eq:DGL_rho}
one has to use a finite basis of Floquet states.
The validity of this approximation is discussed in Ref.~\cite{HonKetKoh2009}.

\section{The switching mechanism}\label{sec:switching}

We will demonstrate the switching process by studying
the asymptotic spatial probability density
averaged over one period of the driving
\begin{equation}\label{eq:rho_x}
\rho(x) := \lim_{t \to \infty} \frac{1}{T} \int_t^{t+T} \diff t'\, \left\langle x \right| \rho(t') \left| x \right\rangle
.
\end {equation}
It can be expressed in terms of the 
solutions $\rho_{ij}$ of Eq.~\eqref{eq:DGL_rho} by
$\rho(x) = \sum_{i,j} \rho_{ij} \frac{1}{T} \int_0^{T} \diff t\; u^*_j(x,t) u_i(x,t)$.
Figure~\ref{fig:bild1}(b) shows that for
the undriven double well, $A = 0$, in thermodynamic equilibrium
at low temperatures
almost all probability is in the left well.
This reflects the dominant occupation of the ground state.
Figure~\ref{fig:bild1}(c) shows that for a small driving amplitude,
$A_0 \approx 0.008$, the probability density is almost
completely transferred to the right well.
Note, that the driving amplitude is so small, that at all times
the right well is energetically higher than the left well.
This example demonstrates that a weak periodic driving not only alters
the static Boltzmann occupation probabilities
\cite{KohDitHae1997, HonKetKoh2009}, but can switch to an
almost orthogonal and macroscopically different asymptotic state
of the system.

\begin{figure}[t]
\includegraphics[width=\columnwidth]{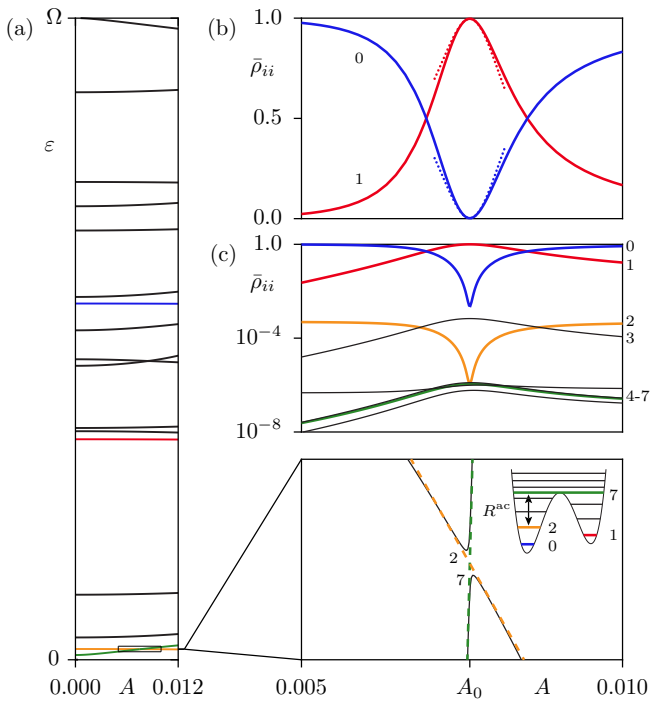}
\caption{
a) Quasienergy spectrum for the 17 lowest Floquet states vs. driving strength $A$
and magnification of the avoided crossing at $A = A_0$ (solid lines) with 
$\Delta = \left|\varepsilon_2(A_0) - \varepsilon_7(A_0) \right| \approx 1.82 \cdot 10^{-6}$,
the quasienergies corresponding to diabatic states 2 and 7 (dashed lines),
and eigenenergies of the undriven potential (inset).
b) Stationary occupations $\bar \rho_{ii}$ in the diabatic basis (solid lines)
and approximation based on effective rate $\Rac$,
Eqs.~\eqref{eq:LGS_rho_AC} and \eqref{eq:Rac} (dotted lines).
c) same as b) with logarithmic axis for $\bar \rho_{ii}$.
The parameters are $\mu = 0.03$, $\heff = 0.04$, $\Omega=\heff/0.768$,
$\beta = 100$, $\eta = 10^{-4}$ and $\omega_c = 100$.
}
\label{fig:bild2}
\end{figure}

We get a first insight into this dramatic phenomenon from Figs.~\ref{fig:bild2}(a)
and (b),
where one can see that under the variation of the driving amplitude $A$
the quasienergy spectrum shows around $A = A_0$
an isolated avoided crossing of the states 2 and 7
originating from the second and the 7th excited state of
the undriven system. We emphasize, that both the ground state,
which is dominantly populated at $A=0$, and the first excited state, which
will turn out to be dominantly populated at $A=A_0$, are not involved in this
avoided crossing.

An intuitive understanding of the switching from the rate equation
seems impossible:
Tuning through an avoided crossing of the two Floquet states 2 and 7
they exchange their character and thus
drastically affect in Eq.~\eqref{eq:DGL_rho} a large number of rates
$R_{ij;kl}$, where one of the four indices is 2 or 7.
In order to visualize the changes of the density operator due to
this avoided crossing it is convenient to express this operator
in a basis that does not significantly change in the neighborhood
of the avoided crossing.
In the subspace of the Floquet states of the avoided crossing
we use the diabatic states 2 and 7,
which would correspond to an exact crossing.
Due to the weak driving amplitude $A \ll \mu$ they are nearly
identical to the eigenstates of the undriven system
(Fig.~\ref{fig:bild2}(b), inset).
We will denote quantities in this diabatic basis by a bar.

The diagonal density matrix elements
$\bar \rho_{ii}$ in the diabatic basis
are shown in Fig.~\ref{fig:bild2}(b) and (c).
One observes that $\bar \rho_{00}$, which corresponds to being in the
ground state of the undriven system, drops from close to one to almost zero
for $A = A_0$.
In contrast, the probability $\bar \rho_{11}$ increases almost to one,
which corresponds to the first excited state
being dominantly populated.
The tiny occupations $\bar \rho_{22}$ and  $\bar \rho_{77}$,
i.e.\ the probabilities to be in one of the states of the avoided crossing,
become equal.
These observations for $\bar \rho_{ii}$ are consistent with the
spatial probability density observed in Fig.~\ref{fig:bild1}(c) and
can indeed be exploited for a switching between the wells:
Tuning the driving amplitude from outside the avoided crossing
into its center is accompanied by a probability transfer from
the former ground state in the left well to the first excited state
localized in the right well.

While the equality $\bar \rho_{22} \simeq \bar \rho_{77}$
at the center of an avoided crossing of states 2 and 7 is quite plausible,
the main question is still unanswered:
How can states 0 and 1, which are not involved in the avoided crossing,
interchange their probability?

\subsection{Effective rate equations}\label{sec:effratesystem}

We will answer the above question by using an effective approximate
rate system introduced in Ref.~\cite{HonKetKoh2009}, which is
derived from Eq.~\eqref{eq:DGL_rho},
\begin{equation}\label{eq:LGS_rho_AC}
0 = - \bar \rho_{ii} \sum_k \bar R_{ik}
    + \sum_k \bar \rho_{kk} \bar R_{ki}  ,
\end{equation}
for the diagonal elements $\bar \rho_{ii}$ in the diabatic basis
with an additional rate
\begin{equation}\label{eq:Rac}
\Rac := \frac{\Gamma}{\left(\Gamma / \Delta\right)^2 + 4 d^2}
\end{equation}
replacing the rates
$\bar R_{27}$, $\bar R_{72}$
in Eq.~\eqref{eq:LGS_rho_AC}
due to the single isolated avoided crossing of diabatic states 2 and 7.
Before we make use of these equations, we make a number of remarks:
The rates $\bar R_{ik} \equiv \bar R_{ik;ik}$ are
expressed in the diabatic basis.
The rate
$\Gamma = \Gamma_2 + \Gamma_7 + \bar R_{22} + \bar R_{77} - 2\bar R_{22;77}$
with $\Gamma_i = \sum_{k \neq i} \bar R_{ik}$ ($i=2,7$)
describes the transitions from the states of the
avoided crossing to all other states.
It is proportional to $\eta$ with a factor that is specific to an
individual avoided crossing.
The rate $\Rac$ depends on
the minimal splitting $\Delta$ of the avoided crossing
and the dimensionless distance
$d := (\bar \varepsilon_7 - \bar \varepsilon_2)/\Delta$
from the avoided crossing.
The main assumptions used in the derivation~\cite{HonKetKoh2009} is that
all quasienergy splittings $\varepsilon_{ij}$, apart from the
isolated avoided crossing of interest, are much larger than the
rates $R_{ij;kl}$. This is fulfilled for a sufficiently weak coupling to
the heat bath and allows for neglecting almost all off-diagonal
density matrix elements.
The only non-negligible off-diagonal elements are
$\bar\rho_{27}$ and $\bar\rho_{72}$, which are decoupled from
Eq.~\eqref{eq:LGS_rho_AC} and proportional to
$\bar\rho_{22} - \bar\rho_{77}$.
The dotted lines in Figs.~\ref{fig:bild2}(b) demonstrate this
approximation.

The main advantage of the effective rate system
in the diabatic basis, Eq.~\eqref{eq:LGS_rho_AC},
is, that tuning the distance $d$ from the avoided crossing affects exclusively
the rate $\Rac$.
In the center of the avoided crossing, $d=0$,
and for a small enough coupling to
the heat bath, $\Gamma \ll \Delta$, it is much larger than all other
rates. This leads directly to almost equal occupations
of the diabatic states involved in the avoided crossing,
$\bar \rho_{22} \simeq \bar \rho_{77}$.
We explain the dominant occupation of state 1
as the combined result of the following facts:
(i)
The overall stationary probability flux between
any two states is in general nonzero,
as detailed balance is broken by the periodic driving.
(ii)
The rates between neighboring states localized in the same well
are much larger than other intra-well rates
as well as inter-well rates
(due to the small spatial overlap between the states of different wells).
Therefore, among the states confined to the same well
detailed balance approximately holds true,
e.g.\ between state 0 and 2 or state 1 and 3.
(iii)
$\Rac$ is the dominant rate and induces occupation equality
of states 2 and 7.
Figure \ref{fig:bild2}(c) shows the resulting  depopulation  of
state 2 towards state 7.
(iv)
Due to the approximate detailed balance among the states in the left well
the relative occupation of state 0 and 2 remains
constant and therefore $\bar \rho_{00}$ drops down together
with $\bar \rho_{22}$.
(v)
The states in the right well equilibrate as before but with increased
weights due to probability conservation.
This explains the switching process observed in Fig.~\ref{fig:bild1}.

\begin{figure}[t]
\includegraphics[width=\columnwidth]{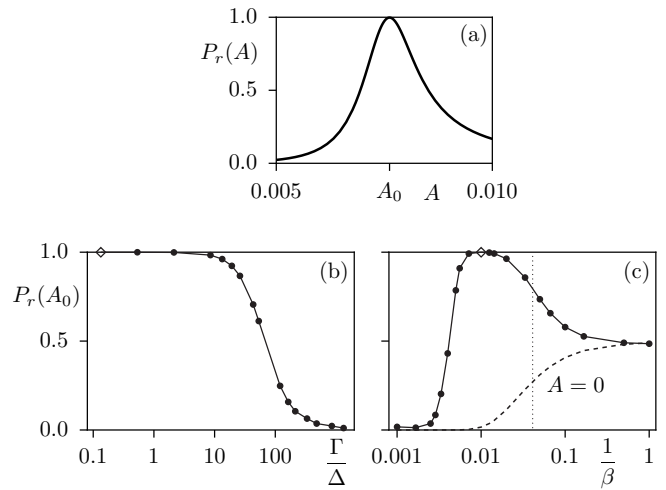}
\caption{
a) Total probability in the right well $P_r$
vs. driving amplitude $A$.
b), c) Peak height $P_r(A_0)$
vs. effective coupling strength $\Gamma/\Delta$ and temperature $1/\beta$.
Diamonds indicate the parameters of Fig.~\ref{fig:bild2}.
The dashed line in c) gives the probability in the right well
without driving, $A = 0$.
The dotted line at $1/\beta = E_1-E_0$
indicates the transition between the high and the low-temperature regimes.
}
\label{fig:bild3}
\end{figure}

An additional surprising phenomenon is observed in Fig.~\ref{fig:bild2}(b).
The impact of the avoided crossing on the occupations
occurs within a significantly broader range 
of the driving amplitude $A$ compared to the width of the avoided crossing.
For the parameters of Fig.~\ref{fig:bild2} the full width at half maximum
of $P_r(A)=\int_0^\infty \diff x \rho(x)$,
the probability to be in the right potential well (Fig.~\ref{fig:bild3}(a)),
is a factor of 30 larger than the width of the avoided crossing.
According to Eqs.~\eqref{eq:LGS_rho_AC} and \eqref{eq:Rac},
the occupations change, if the magnitude of $\Rac$ is larger than
or comparable to other significant rates in Eq.~\eqref{eq:LGS_rho_AC}.
Since these rates vary over many orders of magnitude,
this criterion may be fulfilled even beyond the avoided crossing,
$\left|d\right| > 1$, qualitatively explaining the
enlarged width of $P_r(A)$.

\subsection{Parameter dependence}\label{sec:parameter}

What are the optimal parameters for this switching effect?
A maximal switching efficiency is achieved by a high value of
the probability in the right well
$P_r(A)=\int_0^\infty \diff x \rho(x)$.
This quantity is shown in Fig.~\ref{fig:bild3}(a).
Figure~\ref{fig:bild3}(b) demonstrates that if the
coupling to the heat bath is larger than the
minimal splitting of the avoided crossing, $\Gamma > 100 \Delta$,
almost no probability is switched to the right well.
(Note, that even for the largest values of $\Gamma$ in Fig.~\ref{fig:bild3}(b)
the assumption of weak coupling 
of the Floquet-Markov approach is still fulfilled.)
This is due to the fact that in this limit
$\Rac$ becomes negligible compared to the other rates and thus
the influence of the avoided crossing vanishes~\cite{HonKetKoh2009}.
In contrast, for small coupling $\Gamma < \Delta$ we have a high switching
efficiency and one can show that it is independent
of $\Gamma$ in the limit $\Gamma \to 0$.

Figure~\ref{fig:bild3}(c) shows the influence of the temperature $1/\beta$,
which can be related to the level spacing $E_1-E_0$ of the undriven system.
At high temperatures, $1/\beta \gg E_1-E_0$,
the Floquet states are almost equally
occupied resulting in $P_r(A_0) \approx 0.5$.
For temperatures $1/\beta < E_1-E_0$
the probability in the right well becomes dominant
(while, of course, it vanishes in the undriven case $A = 0$).
For even lower temperatures, however, $P_r(A_0)$ drops to zero.
Here, the occupation equality $\bar \rho_{22} \simeq \bar \rho_{77}$
is rendered by an increase of $\bar \rho_{77}$ towards $\bar \rho_{22}$,
in contrast to the decrease of $\bar \rho_{22}$ towards $\bar \rho_{77}$
in Fig.~\ref{fig:bild2}(c).
Together with $\bar \rho_{22}$ also $\bar \rho_{00}$ remains constant
and therefore switching does not take place.
The origin of this low-temperature dependence remains open.

\subsection{Minimal example}\label{sec:minimal}

\begin{figure}[tbh]
\includegraphics[width=\columnwidth]{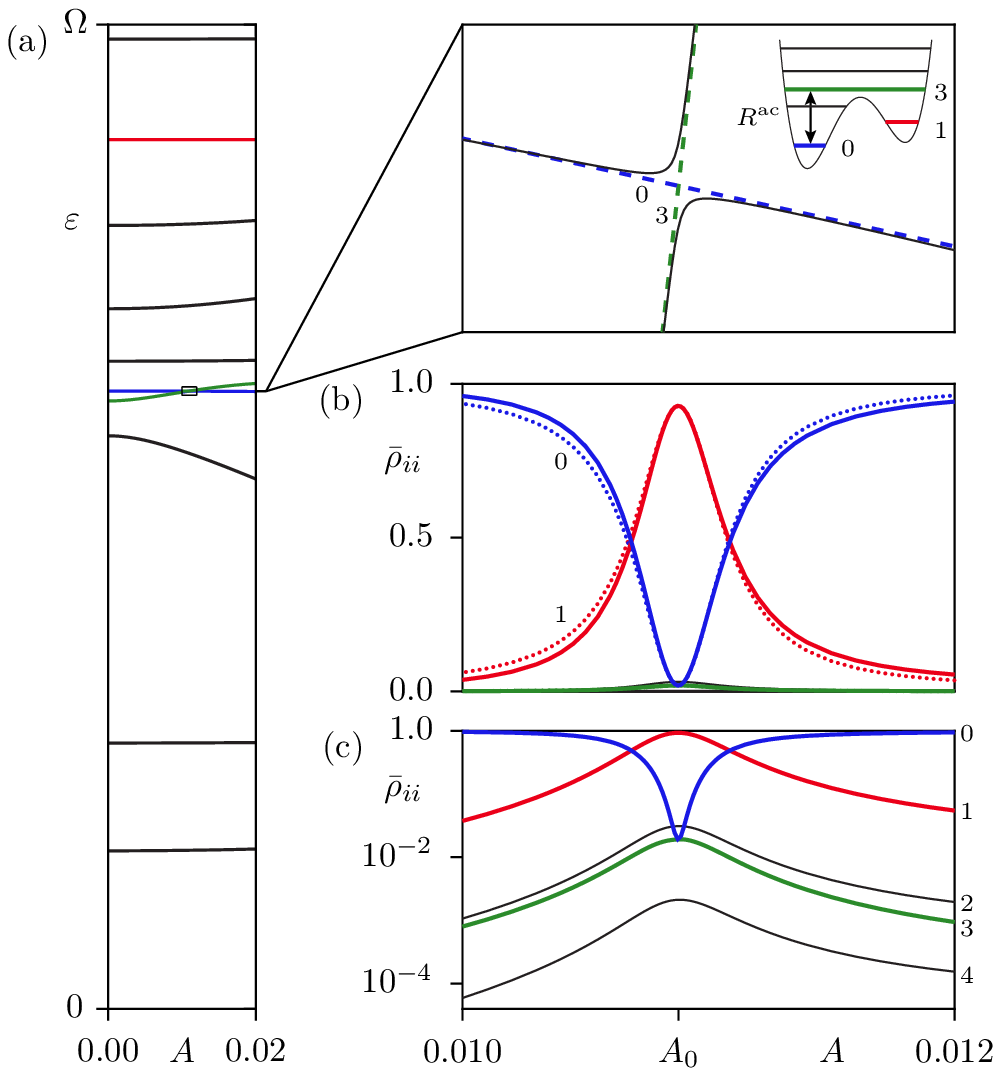}
\caption{
a) Quasienergy spectrum for the ten lowest Floquet states vs. driving strength $A$
and magnification of the avoided crossing at $A = A_0$ (solid lines) with 
$\Delta = \left|\varepsilon_0(A_0) - \varepsilon_3(A_0) \right|\approx 3.57 \cdot 10^{-6}$,
the quasienergies corresponding to diabatic states 0 and 3 (dashed lines),
and eigenenergies of the undriven potential (inset).
b) Stationary occupations $\bar \rho_{ii}$ in the diabatic basis (solid lines)
and approximation based on effective rate $\Rac$,
Eqs.~\eqref{eq:LGS_rho_AC} and \eqref{eq:Rac} (dotted lines).
c) same as b) with logarithmic axis for $\bar \rho_{ii}$.
The parameters are $\mu = 0.08$, $\heff = 0.1$,
$\Omega=0.08165$, $\beta = 60$, $\eta = 10^{-6}$ and $\omega_c = 100$.
}
\label{fig:bild4}
\end{figure}

\begin{figure}[tbh]
\includegraphics[width=\columnwidth]{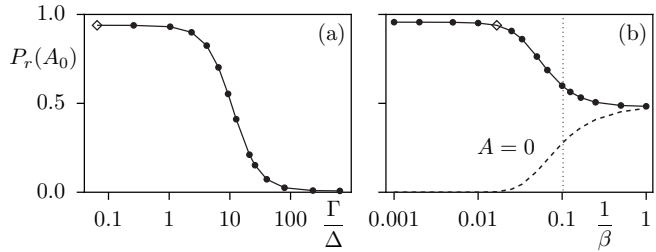}
\caption{
a), b) Peak height $P_r(A_0)$ of the total probability in the right well
vs. effective coupling strength $\Gamma/\Delta$ and temperature $1/\beta$.
Diamonds indicate the parameters of Fig.~\ref{fig:bild4}.
The dashed line in b) gives the probability in the right well
without driving, $A = 0$.
The dotted line at $1/\beta = E_1-E_0$ indicates
the transition between the high and the low-temperature regimes.
}
\label{fig:bild5}
\end{figure}

A minimal example, where one of the partners of the avoided crossing is the
ground state in the left well, is shown in Fig.~\ref{fig:bild4}.
The above discussion then simplifies since step (iv) is eliminated
and essentially just three states are involved.
The parameter dependence of $P_r(A_0)$ on the effective coupling strength
$\Gamma/\Delta$ is unchanged
and the switching mechanism is maintained even for low temperatures,
see Fig.~\ref{fig:bild5}.

For a clear presentation we have above chosen examples in the limit of a
small driving amplitude, where the Floquet states are not very different from
the eigenstates of the undriven system. In this case an avoided crossing
requires near-resonant driving, $E_7-E_2 \approx 3 \Omega$
in Fig.~\ref{fig:bild2} and $E_3-E_0 \approx 3 \Omega$ in Fig.~\ref{fig:bild4}.
We have observed switching also in the case of strong driving,
supporting the generality of the proposed switching mechanism.

\section{Conclusion}\label{sec:conclusion}

In conclusion, we demonstrate a new switching mechanism
for an asymmetric double well potential under
a weak periodic driving and a weak coupling to a heat bath.
As the origin of the switching we identify an
avoided crossing in the quasienergy spectrum of the system.
Under its influence the asymptotic occupations 
of all Floquet states dramatically change
even if both involved Floquet states have just small occupations.
We explain this switching mechanism 
by an effective rate equation at the avoided crossing.

We now briefly discuss possible advantages of the switching mechanism
in applications:
(i) If one uses a laser for the periodic driving, the amplitude dependence of
the switching mechanism and the beam profile allow switching at a
3D spatially localized position with a resolution
smaller than the focus width.
(ii) In situations where a theoretical modeling of the system,
e.g.\ a complex molecule, is not achievable and
no other switching mechanism is known,
the generic appearance of avoided crossings in time-periodically driven systems
suggests the existence of driving parameters for the desired switching.

We emphasize that this switching mechanism is completely different from
standard techniques which allow to transfer a wave packet from one well
to the other by resonant or near resonant driving and
negligible coupling to a heat bath.
There one has to prepare a specific initial wave packet
and has to apply the driving for a specific duration.
In contrast, here the initial state of the system is arbitrary,
the duration of the driving is arbitrary
(if larger than the relaxation time),
and the presence of the heat bath is essential.

\section*{Acknowledgements}

We acknowledge helpful discussions with D.~Hone, S.~Kohler,
and W.~Kohn.
R.K.~thanks the Kavli Institute for Theoretical Physics at
UCSB (NSF Grant No. PHY05-51164).

\end{document}